\documentstyle[11pt]{article}

\def\noi{\noindent}

\def\nqq{\hspace{-2em}}

\def\barr{\left(\begin{array}}
\def\earr{\end{array}\right)}
\def\beq{\begin{equation}}
\def\eeq{\end{equation}}
\def\ber{\begin{eqnarray} &&\nqq}
\def\eer{\end{eqnarray}}
\def\eern{\nonumber \end{eqnarray}}
\def\nn{\nonumber\\ &&\nqq}
\def\mm{\\ &&\nqq}

\catcode`\@=11 \@addtoreset{equation}{section}\catcode`\@=12

\newcommand{\diag}{\mathop{\rm diag}\nolimits}

\newcommand{\sh}{\mathop{\rm sh}\nolimits}
\newcommand{\ch}{\mathop{\rm ch}\nolimits}
\newcommand{\btd}{\bigtriangledown}
\newcommand{\btu}{\bigtriangleup}
\newcommand{\eps}{\varepsilon}
\newcommand{\e}[1]{\mathop{\rm e}\nolimits^{#1}}
\newcommand{\half}{\frac{1}{2}}
 
\newcommand{\p}{\partial}
\newcommand{\ts}{\textstyle}

\newcommand{\im}{{\rm i}}

\newcommand{\fnm}{\footnotemark}
\newcommand{\fnt}{\footnotetext}

\begin{document}


\begin{center}
\large\bf
INTEGRABLE MULTIDIMENSIONAL CLASSICAL AND QUANTUM COSMOLOGY FOR
INTERSECTING P-BRANES \\[10pt]

\small\bf
M.A. Grebeniuk\fnm[1]\fnt[1]{mag@gravi.phys.msu.su} \\[5pt]

\it
Moscow State University, Physical Faculty,
Department of Theoretical Physics, Moscow 117234, Russia \\[10pt]

\bf
V.D. Ivashchuk\fnm[2]\fnt[2]{melnikov@fund.phys.msu.su}
and V.N. Melnikov\fnm[2] \\[5pt]

\it
Center for Gravitation and Fundamental Metrology,
VNIIMS, Moscow 117313, Russia
\end{center}

\vspace{10pt}

\small\noi      
Multidimensional cosmological model describing the evolution of one
Einstein space of non-zero curvature and $n$ Ricci-flat internal spaces
is considered. The action contains several dilatonic scalar fields
$\varphi^I$ and antisymmetric forms $A^I$. When forms are chosen to be
proportional of volume forms of $p$-brane submanifolds of internal
space manifold, the Toda-like Lagrange representation is obtained.
Wheeler--De Witt equation for the model is  presented.
The exact solutions in classical and quantum cases 
are obtained when dimensions of $p$-branes and
dilatonic couplings obey some orthogonality conditions. 

\normalsize

\section{Introduction}

In this paper we continue our investigations of multidimensional
gravitational model governed by the action containing several dilatonic
scalar fields and antisymmetric forms \cite{IM}. The action is presented
below (see (\ref{2.1})). Such form of action is typical for special
sectors of supergravitational models \cite{CJS,SS} and may be of
interest when  dealing with superstring and M-theories
\cite{GSW,HTW,D,S}.

Here we consider a cosmological sector of the model from \cite{IM}.
We recall that this  model treats generalized
intersecting $p$-brane solutions. Using the $\sigma$-model
representation of \cite{IM} we reduce the equations of motion to
the pseudo-Euclidean Toda-like Lagrange system \cite{IM3} with
zero-energy constraint. After  separating  one-dimensional
Liouville subsystem corresponding to negative mode (logarithm of
quasivolume \cite{IM3}) we are led to Euclidean Toda-like system
We consider the simplest case of orthogonal vectors in exponents
of the Toda potential and obtain  exact solutions. In this case
we deal with $a_1+\dots+a_1$ Euclidean Toda lattice (sum of $n$
Liouville actions). Recently analogous reduction for forms of equal
rank was done in \cite{LPX}.

In this paper we  consider also  quantum aspects of the model.
Using the $\sigma$-model representation and the standart prescriptions
of quantization we are led to the 
multidimensional Wheeler--De Witt equation.
This equation is solved in "orthogonal"  case.

\section{The model}

Here like in \cite{IM} we consider the model governed by the action
\ber
\label{2.1}          
S=\frac{1}{2\kappa^{2}}\int_{M}d^{D}z\sqrt{|g|}\biggl\{{R}[g]-
2\Lambda-\sum_{I\in\Omega}\Bigl[g^{MN}\partial_{M}\varphi^I
\partial_{N}\varphi^I+\frac{1}{n_I!}\exp(2\lambda_{JI}\varphi^J)
(F^I)^2_g\Bigr]\biggr\}+S_{GH},
\eer 
where $g=g_{MN}dz^{M}\otimes dz^{N}$ is the metric, $\varphi^I$
is a dilatonic scalar field,
\beq             
\label{2.2}
F^I=dA^I=\frac{1}{n_I!} F^I_{M_1 \ldots M_{n_I}}
dz^{M_1}\wedge\ldots\wedge dz^{M_{n_I}}
\eeq
is $n_I$-form ($n_I\ge2$) on $D$-dimensional manifold $M$, $\Lambda$
is a cosmological constant and $\lambda_{JI}\in{\bf R}$, $I,J\in\Omega$.
In (\ref{2.1}) we denote $|g|=|\det(g_{MN})|$,
\beq
\label{2.3}
(F^I)^2_g=F^I_{M_1\ldots M_{n_I}}F^I_{N_1\ldots N_{n_I}}
g^{M_1 N_1}\ldots g^{M_{n_I}N_{n_I}},
\eeq
and $S_{\rm GH}$ is the standard Gibbons-Hawking boundary
term \cite{GH}. This term is essential for a quantum treatment of the
problem. Here $\Omega$  is a non-empty finite set. (The action (\ref{2.1})
with $\Lambda=0$ and equal $n_I$ was considered recently in \cite{KKLP}.
For supergravity models with different $n_I$ see also \cite{LPS1}).

The equations of motion corresponding to (\ref{2.1}) have the following 
form
\ber
\label{2.4}                                                
R_{MN}-\frac{1}{2}g_{MN}R=T_{MN}-\Lambda g_{MN}, \quad
T_{MN}=\sum_{I\in\Omega}\left[T_{MN}^{\varphi^I}+
\exp(2\lambda_{JI}\varphi^J)T_{MN}^{F^I}\right], \mm
\label{2.5}
{\btu}[g]\varphi^J-\sum_{I\in\Omega}\frac{\lambda_{JI}}{n_I!}
\exp\left(2\lambda_{KI}\varphi^K\right)(F^I)^2_g = 0, \quad
{\btd}_{M_1}[g]\left(\exp(2 \lambda_{KI} \varphi^K)
F^{I,M_1\ldots M_{n_I}}\right)=0,
\eer
where                 
\ber
\label{2.6}
T_{MN}^{\varphi^I}=\p_{M}\varphi^I\p_{N}\varphi^I-
\frac{1}{2}g_{MN}\p_{P}\varphi^I\p^{P}\varphi^I, \mm
\label{2.7}
T_{MN}^{F^I}=\frac{1}{n_{I}!}\biggl[-\frac{1}{2}g_{MN}(F^{I})^{2}_{g}+
n_{I}F^{I}_{MM_2\ldots M_{n_I}}F_{N}^{I,M_2\ldots M_{n_I}}\biggr],
\eer
$I,J \in \Omega$.
In (\ref{2.5}) ${\btu}[g]$ and ${\btd}[g]$ are Laplace-Beltrami and
covariant derivative operators respectively corresponding to $g$.

Here we consider the manifold $M={\bf R}\times M_{0}\times\ldots\times M_{n}$,
with the metric
\beq
\label{2.8}
g=w\e{2{\gamma}(u)}du\otimes du+\sum_{i=0}^{n}\e{2\phi^i(u)}g^i,
\eeq
where $w=\pm1$, $u$ is a time variable and 
$g^i=g_{m_{i}n_{i}}(y_i)dy_i^{m_{i}} \otimes dy_i^{n_{i}}$ is a metric on 
$M_{i}$ satisfying the equation 
$R_{m_{i}n_{i}}[g^i]=\lambda_{i}g^i_{m_{i}n_{i}}$,
$m_{i},n_{i}=1,\ldots,d_{i}$; $\lambda_{i}={\rm const}$, $i=0,\ldots,n$.
The functions $\gamma,\phi^{i}:{\bf R_\bullet}\rightarrow{\bf R}$
(${\bf R_\bullet}$ is an open subset of ${\bf R}$) are smooth.

We claim any manifold $M_i$ to be oriented and connected, $i=0,\ldots,n$.
Then the volume $d_i$-form
\beq
\label{2.9}
\tau_i=\sqrt{|g^i(y_i)|}\ dy_i^{1}\wedge\ldots\wedge dy_i^{d_i},
\eeq
and signature parameter $\eps(i)={\rm sign}(\det(g^i_{m_{i}n_{i}}))=\pm1$
are correctly defined for all $i=0,\ldots,n$.

Let $\Omega$ from (\ref{2.1}) be a set of all non-empty subsets of
$\{0,\ldots,n\}$. The number of elements in $\Omega$ is $|\Omega|=2^{n+1}-1$.
For any $I=\{i_1,\ldots,i_k\}\in\Omega$, $i_1<\ldots<i_k$, we put in
(\ref{2.2})
\beq
\label{2.10}
A^I=\Phi^I\tau_{i_1}\wedge\ldots\wedge\tau_{i_k},
\eeq
where functions $\Phi^I:{\bf R_\bullet}\rightarrow{\bf R}$ are smooth,
and $\tau_{i}$ are defined in (\ref{2.9}). In components
relation (\ref{2.10}) reads
\ber
\label{2.11}
A^{I}_{P_1\ldots P_{d(I)}}(u,y)=\Phi^{I}(u)\sqrt{|g^{i_1}(y_{i_1})|}
\ldots\sqrt{|g^{i_k}(y_{i_k})|}\ \eps_{P_1\ldots P_{d(I)}},
\eer
where $d(I)\equiv d_{i_1}+\ldots+d_{i_k}=\sum_{i\in I}d_i$ is 
the dimension
of the oriented manifold $M_{I}=M_{i_1}\times\ldots\times M_{i_k}$, and
indices $P_1,\ldots,P_{d(I)}$ correspond to $M_I$. It follows from
(\ref{2.10}) that
\beq
\label{2.12}
F^I=dA^I=d\Phi^I\wedge\tau_{i_1}\wedge\ldots\wedge\tau_{i_k},
\eeq
or, in components,
\ber
\label{2.13}
F^{I}_{uP_1\ldots P_{d(I)}}=-F^{I}_{P_1u\ldots P_{d(I)}}=
\ldots=\dot\Phi^{I}\sqrt{|g^{i_1}|}\ldots\sqrt{|g^{i_k}|}\
\eps_{P_1\ldots P_{d(I)}},
\eer
and $n_I=d(I)+1$, $I\in\Omega$.

Thus dimensions of forms $F^I$ in the considered model are fixed by the
subsequent decomposition of the manifold.

\section{$\sigma$-model representation}

For dilatonic scalar fields we put $\varphi^I=\varphi^I(u)$. Let
\beq
\label{3.1}
f=\gamma_0-\gamma, \quad \sum_{i=0}^{n}d_i\phi^i\equiv\gamma_0.
\eeq

It is not difficult to verify that the field equations
(\ref{2.4})--(\ref{2.5}) for the field configurations from (\ref{2.8}),
(\ref{2.10}) may be obtained as the equations of motion corresponding
to the action
\ber
\label{3.2}
S_{\sigma}=\frac{1}{2\kappa^{2}_0}\int du\e{f}\biggl\{-wG_{ij}\dot
\phi^i\dot\phi^j-w\delta_{IJ}\dot\varphi^I\dot\varphi^J \nn
-w\sum_{I\in\Omega}\eps(I)\exp\Bigl(2\vec\lambda_I\vec\varphi-
2\sum_{i\in I}d_i\phi^i\Bigr)(\dot\Phi^I)^2-2V\e{-2f}\biggr\},
\eer
where $\vec\varphi=(\varphi^I)$, $\vec\lambda_I=(\lambda_{JI})$,
$\dot\varphi\equiv d\varphi(u)/du$; $G_{ij}=d_i\delta_{ij}-d_id_j$,
are component of "pure cosmological" minisuperspace metric and
\beq
\label{3.3}
V={V}(\phi)=\Lambda\e{2{\gamma_0}(\phi)}-\half\sum_{i =1}^{n}
\lambda_id_i\e{-2\phi^i+2{\gamma_0}(\phi)}
\eeq
is the potential. In (\ref{3.2}) $\eps(I)\equiv\eps(i_1)\times
\ldots\times\eps(i_k)=\pm1$ for $I= \{i_1,\ldots,i_k \} \in\Omega$.
   
For finite internal space volumes $V_i$ (e.g. compact $M_i$)      
the action (\ref{3.2}) coincides with the action (\ref{2.1}) if
$\kappa^{2}=\kappa^{2}_0\prod_{i=0}^{n}V_i$.

The representation (\ref{3.2}) follows from more general $\sigma$-model
action from \cite{IM4}, that may be written in the following form
\ber
\label{3.4}
S_\sigma=\frac\mu2\int du\Bigl\{(-w){\cal N}{\cal G}_{\hat A\hat B}(\sigma)
\sigma^{\hat A}\sigma^{\hat B}-2{\cal N}^{-1}V(\sigma)\Bigr\},
\eer
where $(\sigma^{\hat A})=(\phi^i,\varphi^I,\Phi^{I'})\in{\bf R}^{n+1+2m}$,
$m=|\Omega|$, $\mu=1/(2k_0^2)$; ${\cal N}=\exp(\gamma_0-\gamma)>0$ is 
the lapse function and
\ber
\label{3.5}
({\cal G}_{\hat A\hat B})=\left(\begin{array}{ccc}
G_{ij}&0&0\\[5pt]
0&\delta_{IJ}&0\\[5pt]
0&0&\eps(I')\exp\left(2\vec\lambda_{I'}
\vec\varphi-2\sum\limits_{i\in I'}d_i\phi^i\right)\delta_{I'J'}
\end{array}\right)
\eer
is matrix of minisupermetric of the model (target space metric),
$i,j=0,\dots,n$; $I,J,I',J'\in\Omega$.

Let us fix the gauge in (\ref{3.1}):  $f = f(\sigma)$, where 
$f(\sigma)$ is smooth. We call this gauge as $f$-gauge.
>From (\ref{3.4}) we get the Lagrange system with the Lagrangian and
the energy constraint
\ber
\label{3.6}
L=\frac\mu2\e{f}{\cal G}_{\hat A\hat B}(\sigma)
\sigma^{\hat A}\sigma^{\hat B}+w\mu\e{-f}V(\sigma), \mm
\label{3.7}
E=\frac\mu2\e{f}{\cal G}_{\hat A\hat B}(\sigma)
\sigma^{\hat A}\sigma^{\hat B}-w\mu\e{-f}V(\sigma)=0.
\eer
We note that  the minisupermetric
${\cal G}=
{\cal G}_{\hat A\hat B}d\sigma^{\hat A}\otimes d\sigma^{\hat B}$
is not flat. Here the problem of integrability of Lagrange
equations for the Lagrangian (\ref{3.6}) arises.

The minisuperspace metric (\ref{3.5}) may be also written in the form
\beq
\label{3.8}
{\cal G}=\bar G+\sum_{I\in\Omega}\eps(I)\e{-2U^I(x)}
d\Phi^I\otimes d\Phi^I,
\quad
U^I(x)=\sum_{i\in I}d_i\phi^i-\vec\lambda_I\vec\varphi,
\eeq
where $x=(x^A)=(\phi^i,\varphi^I)$, $\bar G=\bar G_{AB}dx^A\otimes dx^B=
G_{ij}d\phi^i\otimes d\phi^j+\delta_{IJ}d\varphi^I\otimes d\varphi^J$,
\ber
\label{3.9}
(\bar G_{AB})=\left(\begin{array}{cc}
G_{ij}&0\\
0&\delta_{IJ}
\end{array}\right)
\eer
$i,j\in\{0,\dots,n\}$, $I,J\in\Omega$, and the potential $(-w)V$ may
be presented in the following form
\ber
\label{3.10}
(-w)V=\sum_{j=0}^n\left(\frac w2\lambda_id_i\right)\exp[2U^i(x)]+
(-w)\Lambda\exp[2U^\Lambda(x)],
\eer
where
\ber
\label{3.11}
U^\Lambda(x)=U_A^\Lambda x^A=\sum_{j=0}^nd_j\phi^j, \quad
U^i(x)=-\phi^i+\sum_{j=0}^nd_j\phi^j, 
\eer
or in components
\ber
\label{3.12}
(U_A^i)=(-\delta_j^i+d_j,0), \quad
(U_A^\Lambda)=(d_j,0), \quad
(U_A^I)=\left(\sum_{i\in I}\delta_j^id_i,-\lambda_{JI}\right),
\eer
$i,j\in\{0,\dots,n\}$, $I, J\in\Omega$.

Let $(x,y)\equiv\bar G_{AB}x^Ay^B$
define a quadratic form on ${\cal V}= {\bf R}^{n+1+m}$, 
$m=2^{n+1}-1$. 
The dual form defined on dual space ${\cal V}^*$ is following
\ber
\label{3.13}
(U,U')_*=\bar G^{AB}U_AU'_B,
\eer
where $U,U'\in{\cal V}^*$, $U(x)=U_Ax^A$, $U'(x)=U'_Ax^A$ and $(\bar G^{AB})$
is the  matrix inverse to the matrix $(\bar G_{AB})$. Here, like in 
\cite{IMZ},
\ber
\label{3.14}
G^{ij}=\frac{\delta^{ij}}{d_i}+\frac1{2-D}
\eer
$i,j=0,\dots,n$.

The integrability of the Lagrange system crucially depends upon the scalar
products (\ref{3.13}) for vectors $U^i$, $U^\Lambda$, $U^I$ from 
(\ref{3.8}), (\ref{3.11}).
Here we present these scalar products
\ber
\label{3.15}
(U^i,U^j)_*=\frac{\delta_{ij}}{d_j}-1, \quad
(U^i,U^\Lambda)_*=-1, \quad
(U^\Lambda,U^\Lambda)_*=-\frac{D-1}{D-2}, \mm
\label{3.16}
(U^I,U^i)_*=-\frac{d(I\cap\{i\})}{d_i}, \quad
(U^I,U^\Lambda)_*=\frac{d(I)}{2-D}, \mm
\label{3.17}
(U^I,U^J)_*=q(I,J)+\vec\lambda_I\vec\lambda_J, \quad
q(I,J)\equiv d(I\cap J)+\frac{d(I)d(J)}{2-D},
\eer
$I,J\in\Omega$, $i,j=0,\dots,n$.

The relations (\ref{3.15}) were calculated in  \cite{GIM},
the relations (\ref{3.17}) were obtained in \cite{IM} ($U_A^I=-L_{AI}$
in notations of \cite{IM}).

\section{Classical exact solutions}

Here we will integrate the Lagrange equations corresponding to the
Lagrangian (\ref{3.6}) with the energy-constraint (\ref{3.7}).
We put $f=0$, i.e. the harmonic time gauge is considered.

The problem of integrability may be simplified if we integrate the
Maxwell equations
\ber
\label{5.1}
\frac d{du}\left(\exp\Bigl(2\vec\lambda_I\vec\varphi-
2\sum_{i\in I}d_i\phi^i\Bigr)\dot\Phi^I\right)=0, \quad
\dot\Phi^I=Q^I\exp\left(-2\vec\lambda_I\vec\varphi+
2\sum_{i\in I}d_i\phi^i\right),
\eer
where $Q^I$ are constant, $I\in\Omega$.

Let $Q^I\ne0\Leftrightarrow I\in\Omega_*$, where $\Omega_*\subset\Omega$
is some non-empty subset of $\Omega$. For fixed $Q=(Q^I,I\in\Omega_*)$
the Lagrange equations corresponding to $\phi^i$ and $\varphi^I$, when
equations (\ref{5.1}) are substituted, are equivalent to the Lagrange
equations for the Lagrangian
\ber
\label{5.4}
L_Q=\frac12 \bar G_{AB}\dot x^A\dot x^B-V_Q, 
\eer
where   $x=(x^A)=(\phi^i,\varphi^I)$, $i=0, \ldots, n $, 
$I \in \Omega$ and 
\ber
\label{5.6}
V_Q= (-w)V+\sum_{I\in\Omega_*}\frac12\eps(I)(Q^I)^2\exp[2U^I(x)],
\eer
(for $(-w)V$ see (\ref{3.10})). Thus, we are led to the
pseudo-Euclidean Toda-like system (see \cite{IMZ,IM3}) with the
zero-energy constraint:
\ber
\label{5.4a}
E_Q=\frac12\bar G_{AB}\dot x^A\dot x^B+V_Q=0.
\eer

\subsection{The case $\Lambda=\lambda_i=0$, $i=1,\dots,n$.}

Here we put $\Lambda=0$, $\lambda_i=0$ for $i=1,\dots,n$; and
$\lambda_0\ne0$. In this case the potential (\ref{5.6})
\ber
\label{5.7}
V_Q=\left(\frac w2\lambda_0d_0\right)\exp[2U^0(x)]+
\sum_{I\in\Omega_*}\frac{\eps(I)}2(Q^I)^2\exp[2U^I(x)]
\eer
is governed by time-like vector $U^0$ $(d_0>1)$ and $m_*=|\Omega_*|$
space-like vectors $U^I$:
\ber
\label{5.8}
(U^0,U^0)=\frac1{d_0}-1<0, \quad (U^I,U^I)=
d(I)\frac{D-2-d(I)}{D-2}+(\vec \lambda_I)^2>0
\eer

We also put $0\notin I, \quad \forall I\in\Omega_*$. This condition means
that all $p$-branes do not contain the manifold $M_0$. It follows from
(\ref{3.16}) that
\ber
\label{5.9}
(U^I,U^0)_*=0,
\eer
for all $I\in\Omega_*$. In this case the Lagrangian (\ref{5.4}) with
the potential (\ref{5.7}) may be diagonalized by linear coordinate
transformation $z^a=S_i^ax^i$, $a=0,\dots,n$, satisfying
$\eta_{ab}S_i^aS_j^b=G_{ij}$, where $\eta_{ab}=\diag(-1,+1,\dots,+1)$.
There exists the diagonalization such that $U_i^0x^i=q_0z^0$, where
\ber
\label{5.10}
q_0=\sqrt{-(U^0,U^0)_*}=\sqrt{1-\frac1{d_0}}
\eer
is a parameter, $q_0<1$.

In $z$-variables the Lagrangian (\ref{5.4}) reads $L_Q=L_0+L_Q^E$, where
\ber
\label{5.11}
L_0=-\frac12(\dot z^0)^2-A_0\exp(2q^0z^0), \mm
\label{5.12}
L_Q^E=\frac12\left[(\dot{\vec z})^2+\delta_{IJ}\dot\varphi^I
\dot\varphi^J\right]-\sum_{I\in\Omega_*}A_I\exp\left(2\vec q_I\vec z-
2\lambda_{JI}\varphi^J\right).
\eer
In (\ref{5.11}), (\ref{5.12}) $\vec z=(z^1,\dots,z^{n})$,
$\vec q_I=(q_{I,1},\dots,q_{I,n})$, $A_0\equiv(w/2)\lambda_0d_0$,
$A_I=(1/2)\eps(I)(Q^I)^2$ and $\vec q_I\cdot\vec q_J=q(I,J)$,
$I,J\in\Omega_*$.

Thus the Lagrangian (\ref{5.4}) is splitted into sum of two
independent parts (\ref{5.11}) and (\ref{5.12}). The latter may
be written as
\ber
\label{5.13}
L_Q^E=\frac12(\dot{\vec Z})^2-\sum_{I\in\Omega_*}A_I\exp(2\vec B_I\vec Z)
\eer
where $\vec Z=(z^1,\dots,z^{n},\varphi^I)$, $\vec B_I=(\vec q_I,\lambda_{JI})$.
Thus the equations of motion for the considered cosmological model are
reduced to the equations of motion for the Lagrange systems with the
Lagrangians (\ref{5.11}) and (\ref{5.12}) and the energy constraint
$E=E_0+E_Q^E$ = 0, where
\ber
\label{5.14}
E_0=-\frac12(\dot z^0)^2+A_0\exp(2q_0z^0), \quad
E_Q^E=\frac12(\dot{\vec Z})^2+\sum_{I\in\Omega_*}A_I\exp(2\vec B_I\vec Z).
\eer

The vectors $\vec B_I$ in (\ref{5.13}) satisfy the relations
\ber
\label{5.15}
\vec B_I\cdot\vec B_J=q(I,J)+\vec\lambda_I\vec\lambda_J,
\eer
$I,J\in\Omega_*$, where $p$-brane "overlapping index" $q(I,J)$ is
defined in (\ref{3.17}).

\subsection{The case of orthogonal $\vec B_I$}

The simplest situation arises when the vectors $\vec B_I$ are
orthogonal, i.e.
\ber
\label{5.16}
\vec B_I\vec B_J=(U^I,U^J)_*=d(I\cap J)+\frac{d(I)d(J)}{2-D}+
\vec\lambda_I\vec\lambda_J=0,
\eer
for all $I\ne J$, $I,J\in\Omega_*$. In this case the Lagrangian
(\ref{5.13}) may be splitted into the sum of $|\Omega_*|$
Lagrangians of the Liouville type and $n$ "free" Lagrangians.

Using relations from \cite{GIM} we readily obtain  exact solutions
for the Euler-Lagrange equations corresponding to the Lagrangian
(\ref{5.4}) with the potential (\ref{5.7}) 
 when the orthogonality conditions
(\ref{5.9}) and (\ref{5.16}) are satisfied.

The solutions for  $(x^A)=(\phi^i,\varphi^I)$ read
\ber
\label{5.17}
x^A(u)=-\frac{U^{0A}}{(U^0,U^0)_*}\ln|f_0(u-u_0)|-
\sum_{I\in\Omega_*}\frac{U^{IA}}{(U^I,U^I)_*}\ln|f_I(u-u_I)|+
\alpha^A u+\beta^A,
\eer
where $u_0,u_I$
are constants, $U^{sA}\equiv\bar G^{AB}U_B^s$
are contravariant components of $U^s$, $s\in\{0\}\sqcup\Omega_*$,
$\bar G^{AB}$ is the matrix inverse to the matrix (\ref{3.9}).
Functions $f_0$, $f_I$ in (\ref{5.17}) are the following
\ber
\label{5.18}
f_0(\tau)=
\left|\frac{\lambda_0(d_0-1)}{C_0}\right|^{\ts\frac12}
\sh(\sqrt{C_0}\tau), \; C_0>0, \; \lambda_0w>0; \quad
\left|\frac{\lambda_0(d_0-1)}{C_0}\right|^{\ts\frac12}
\sin(\sqrt{|C_0|}\tau), \; C_0<0, \; \lambda_0w>0; \nn
\phantom{f_0(\tau)=}
\left|\frac{\lambda_0(d_0-1)}{C_0}\right|^{\ts\frac12}
\ch(\sqrt{C_0}\tau), \; C_0>0, \; \lambda_0w<0; \quad
\left|\lambda_0(d_0-1)\right|^{\ts\frac12}
\tau, \; C_0=0, \; \lambda_0w>0;
\eer
and
\ber
\label{5.19}
f_I(\tau)=
\frac{|Q^I|}{\nu_I|C_I|^{1/2}}\sh(\sqrt{C_I}\tau), \;
C_I>0, \; \eps(I)<0; \quad
\frac{|Q^I|}{\nu_I|C_I|^{1/2}}\sin(\sqrt{|C_I|}\tau), \;
C_I<0, \; \eps(I)<0; \nn
\phantom{f_I(\tau)=}
\frac{|Q^I|}{\nu_I|C_I|^{1/2}}\ch(\sqrt{C_I}\tau), \;
C_I>0, \; \eps(I)>0; \quad
\frac{|Q^I|}{\nu_I}\tau, \; C_I=0, \; \eps(I)<0,
\eer
where $C_0$, $C_I$ are constants, and
\ber
\label{5.20}
\nu_I^{-1}=\sqrt{d(I)\left(1+\frac{d(I)}{2-D}\right)+\vec\lambda_I^2}>0,
\eer
$I\in\Omega_*$.

Vectors $\alpha=(\alpha^A)$ and $\beta=(\beta^A)$ in (\ref{5.17})
satisfy the linear constraint relations:
\ber
\label{5.21}
U^0(\alpha)=-\alpha^0+\sum_{j=0}^nd_j\alpha^j=0; \quad
U^0(\beta)=-\beta^0+\sum_{j=0}^nd_j\beta^j=0; \mm
\label{5.22}
U^I(\alpha)=\sum_{i\in I}d_i\alpha^i-\lambda_{JI}\alpha^J=0; \quad
U^I(\beta)=\sum_{i\in I}d_i\beta^i-\lambda_{JI}\beta^J=0.
\eer

Calculations of contravariant components $U^{sA}$ give the
following relations
\ber
\label{5.23}
U^{0i}=-\frac{\delta_0^i}{d_i}, \quad U^{0I}=0, \quad
U^{Ii}= \sum_{k\in I}\delta^{ik}+\frac{d(I)}{2-D},
\quad U^{IJ}=-\lambda_{JI},
\eer
$i=0,\dots,n$; $J,I\in\Omega_*$. Substitution of (\ref{5.23}) and
(\ref{5.8}) into the solution (\ref{5.17}) leads us to the following
relations for the logarithms of scale fields
\ber
\label{5.24}
\phi^i=\frac{\delta_0^i}{1-d_0}\ln|f_0|+\sum_{I\in\Omega_*}
\alpha_I^i\ln|f_I|+\alpha^iu+\beta^i, \quad
\varphi^J=\sum_{I\in\Omega_*}\lambda_{JI}\nu_I^2\ln|f_I|+
\alpha^Ju+\beta^J,
\eer
where $\alpha_I^i=-\left(\sum_{k\in I}\delta^{ik}+d(I)/(2-D)\right)\nu_I^2$,
$i=0,\dots,n$; $I\in\Omega_*$.

For harmonic gauge $\gamma=\gamma_0(\phi)$ we get from (\ref{5.24})
\ber
\label{5.25}
\gamma=\sum_{i=0}^nd_i\phi^i=\frac{d_0}{1-d_0}\ln|f_0|+
\sum_{I\in\Omega_*}\frac{d(I)}{D-2}\nu_I^2\ln|f_I|+
\alpha^0u+\beta^0,
\eer
where $\alpha^0$ and $\beta^0$ are given by (\ref{5.21}).

The zero-energy constraint
\ber
\label{5.26}
E=E_0+\sum_{I\in\Omega_*}E_I+\frac12(\alpha,\alpha)=0,
\eer
where $\alpha=(\alpha^i,\alpha^I)$, $(\alpha,\alpha)=\bar G_{AB}
\alpha^A\alpha^B=G_{ij}\alpha^i\alpha^j+\delta_{IJ}\alpha^I\alpha^J$
and $C_s=2E_s (U^s,U^s)_*$, $s=0,I$, (see \cite{GIM}) may be written
in the following form
\ber
\label{5.27}
C_0\frac{d_0}{d_0-1}=\sum_{I\in\Omega_*}[C_I\nu_I^2+
(\alpha^I)^2]+\sum_{i=1}^nd_i(\alpha^i)^2+
\frac1{d_0-1}\left(\sum_{i=1}^nd_i\alpha^i\right)^2.
\eer

The substitution of relations (\ref{5.24}) into (\ref{5.1}) implies
$\dot\Phi^I=Q_I/f_I^2$ and hence we get for forms
\ber
\label{5.28}
F^I=d\Phi^I\wedge d\tau_I=\frac{Q^I}{f_I^2}du\wedge d\tau_I,
\eer
where $\tau_I\equiv\tau_{i_1}\wedge\dots\wedge\tau_{i_k}$,
$I=\{i_1,\dots,i_k\}\in\Omega_*$, $i_1<\dots<i_k$.

The relation for the metric may be readily obtained using the formulas
(\ref{5.24}), (\ref{5.25}).
\ber
\label{5.29}
g= (\prod_{I\in\Omega_*}[f_I^2(u-u_I)]^{d(I)\nu_I^2/(D-2)})
\Biggl\{[f_0^2(u-u_0)]^{d_0/(1-d_0)}\e{2\alpha^0u+2\beta^0} \nn
\times[wdu\otimes du+f_0^2(u-u_0)g^0]+
\sum_{i\ne0}\biggl(\prod_{I\in\Omega_* \atop I\ni i}
[f_I^2(u-u_I)]^{-\nu_I^2}\biggr)\e{2\alpha^iu+2\beta^i}g^i\Biggr\}.
\eer

\section{Wheeler--De Witt equation}

Let us consider the Lagrangian system with Lagrangian (\ref{3.6}). Using
the standard prescriptions of quantization (see, for example, \cite{IMZ})
we are led to the Wheeler-DeWitt equation
\ber
\label{4.1}
\hat{H}^f\Psi^f\equiv\left(-\frac{1}{2\mu}\Delta\left[e^f{\cal G}\right]+
\frac{a}{\mu}R\left[e^f{\cal G}\right]+e^{-f}\mu (-w)V\right)\Psi^f=0,
\eer
where
\beq
\label{4.2}
a=\frac{N-2}{8(N-1)}, \quad N=n+1+2|\Omega|.
\eeq
Here $\Psi^f=\Psi^f(\sigma)$ is the so-called "wave function of the
universe" corresponding to the $f$-gauge, $\Delta[{\cal G}_1]$
and $R[{\cal G}_1]$ denote the Laplace-Beltrami operator and the scalar
curvature correponding to ${\cal G}_1$. For the scalar
curvature we get
\beq
\label{4.3}
R[{\cal G}]=-\sum_{I\in\Omega}(U^I,U^I)-\sum_{I,I'\in\Omega}(U^I,U^{I'}).
\eeq
For the Laplace operator we obtain
\ber
\label{4.4}
\Delta[{\cal G}]\equiv\frac1{\sqrt{|{\cal G}|}}\partial_{\hat A}
\left({\cal G}^{\hat A\hat B}
\sqrt{|{\cal G}|} \partial_{\hat B}\right)=
\e{U(x)}\frac\partial{\partial x^A}\left(\bar G^{AB}
\e{-U(x)}\frac\partial{\partial x^B}\right)+\sum_{I\in\Omega}\eps(I)
\e{2U^I(x)}\left(\frac\partial{\partial\Phi^I}\right)^2,
\eer
where
\beq
\label{4.5}
U(x)=\sum_{I\in\Omega}U^I(x).
\eeq

{\bf Harmonic-time gauge.} The WDW equation (\ref{4.1}) for $f=0$
\beq
\label{4.6}
\hat H\Psi\equiv\left(-\frac{1}{2\mu}\Delta[{\cal G}]+
\frac{a}{\mu}R[{\cal G}]+\mu (-w) V\right)\Psi=0,
\eeq
may be rewritten, using relations (\ref{4.3}), (\ref{4.4}) as follows
\ber
\label{4.7}
2\mu\hat H\Psi=\Biggl\{-G^{ij}\frac\partial{\partial\phi^i}
\frac\partial{\partial\phi^j}-\delta^{IJ}
\frac\partial{\partial\varphi^I}\frac\partial{\partial\varphi^J}-
\sum_{I\in\Omega}\eps(I)\e{2U^I(x)}
\left(\frac\partial{\partial\Phi^I}\right)^2 \nn
+\sum_{I\in\Omega}\left[\sum_{i\in I}\frac\partial{\partial\phi^i}-
\frac{d(I)}{D-2}\sum_{j=0}^n\frac\partial{\partial\phi^j}-
\lambda_{JI} \frac\partial{\partial\varphi^J}\right]+
2aR[{\cal G}]+ 2 \mu^2 (-w) V\Biggr\}\Psi=0.
\eer
Here $\hat H\equiv\hat H^{f=0}$ and $\Psi=\Psi^{f=0}$.

\section{Quantum solutions}

Let us now consider the solutions of the Wheeler--De Witt equation
(\ref{4.6}) for the case of harmonic-time gauge. We also note that
the orthogonality conditions (\ref{5.9}) and (\ref{5.16}) are satisfied.
In $z$-variables 
$z = (z^A)= (S^A_B x^B) = (z^0, \vec{z}, z^I)$
satisfying $q_0 z^0 = U^0(x)$, $q_I z^I = U^I(x)$,
$q_I=\nu_I^{-1}$, we get
\ber
\label{6.1}
\Delta[{\cal G}]=-\left(\frac\partial{\partial z^0}\right)^2+
\left(\frac\partial{\partial\vec z}\right)^2+
\sum_{I\in\Omega}\e{q_Iz^I}\frac\partial{\partial z^I}
\left(\e{-q_Iz^I}\frac\partial{\partial z^I}\right)+
\sum_{I\in\Omega}\eps(I)\e{2q_Iz^I}
\left(\frac\partial{\partial\Phi^I}\right)^2.
\eer
The relation (\ref{4.3}) in the orthogonal
case reads as
\beq
\label{6.2}
R[{\cal G}]=-2\sum_{I\in\Omega}(U^I,U^I)=-2\sum_{I\in\Omega}q_I^2.
\eeq

We are seeking the solution to WDW equation (\ref{4.6}) by the method
of  separation of variables, i.e. we put
\beq
\label{6.3}
\Psi_*(z)=\Psi_0(z^0) \left(\prod_{I\in\Omega}\Psi_I(z^I) \right)
\e{\im P_I \Phi^I} \e{\im\vec p\vec z}.
\eeq
It follows from (\ref{6.1}) that $\Psi_*(z)$ satisfies the WDW equation
(\ref{4.6}) if
\ber
\label{6.4}
2\hat H_0\Psi_0\equiv\left\{\left(\frac\partial{\partial z^0}\right)^2+
\mu^2w\lambda_0d_0\e{2q_0z^0}\right\}\Psi_0=2{\cal E}_0\Psi_0; \mm
\label{6.5}
2\hat H_I\Psi_I\equiv\left\{-\e{q_Iz^I}\frac\partial{\partial z^I}
\left(\e{-q_Iz^I}\frac\partial{\partial z^I}\right) +
\eps(I)P_I^2\e{2q_Iz^I}\right\}\Psi_I=2{\cal E}_I\Psi_I,
\eer
$I\in\Omega$, and
\beq
\label{6.6}
2{\cal E}_0+(\vec p)^2+2\sum_{I\in\Omega}{\cal E}_I+2aR[{\cal G}]=0,
\eeq
with $a$ and $R[{\cal G}]$ from (\ref{4.2}) and (\ref{6.2}) respectively.

Solving (\ref{6.4}), (\ref{6.5}) we obtain
\ber
\label{6.7}
\Psi_0(z^0)=B_{\omega_0({\cal E}_0)}^0
\left(\mu\sqrt{-w\lambda_0d_0}\ \frac{\e{q_0z^0}}{q_0}\right), \quad
\Psi_I(z^I)=\e{q_Iz^I/2}B_{\omega_I({\cal E}_I)}^I
\left(\sqrt{-\eps(I)P_I^2}\ \frac{\e{q_Iz^I}}{q_I}\right),
\eer
where $\omega_0({\cal E}_0)=\sqrt{2{\cal E}_0}/q_0$, $\omega_I({\cal E}_I)=
\sqrt{1/4-\eps(I)2{\cal E}_I\nu_I^2}$, $I\in\Omega$ $(\nu_I=q_I^{-1})$
and $B_\omega^0,B_\omega^I=I_\omega,K_\omega$ are modified Bessel function.

The general solution of the WDW equation (\ref{4.6}) is a superposition
of the "separated" solutions (\ref{6.3}):
\beq
\label{6.8}
\Psi(z)=\sum_B\int dpdPd{\cal E}C(P,p,{\cal E},B)
\Psi_*(z|P,p,{\cal E},B),
\eeq
where $p=(\vec p)$, $P=(P_I)$, ${\cal E}=({\cal E}_I)$, $B=(B^0,B^I)$,
$B^0,B^I=I,K$, $\Psi_*=\Psi_*(z|P,p,{\cal E},B)$ are given by relation
(\ref{6.3}), (\ref{6.7}) with ${\cal E}_0$ from (\ref{6.6}).

\section{Conclusion}

Thus we obtained exact classical and quantum
solutions for multidimensional cosmology,
describing the evolution of $(n+1)$ spaces $(M_0,g_0),\dots,(M_n,g_n)$,
where $(M_0,g_0)$ is an Einstein space of non-zero curvature, and
$(M_i,g^i)$ are "internal" Ricci-flat spaces, $i=1,\dots,n$; in the
presence of several scalar fields and forms. 

The classical solution is given
by relations (\ref{5.24}), (\ref{5.28}), (\ref{5.29}) with the
functions $f_0$, $f_I$ defined in (\ref{5.18})--(\ref{5.19}) and
the relations on the parameters of solutions $\alpha^s$, $\beta^s$,
$C_s$ $(s=i,I)$, $\nu_I$, (\ref{5.21})--(\ref{5.22}), (\ref{5.20}),
(\ref{5.27}) imposed.
The quantum  solutions are presented
by relations (\ref{6.3}), (\ref{6.6})-(\ref{6.8}).

These solutions  describe a set of charged (by forms) overlapping
$p$-branes "living" on submanifolds  not containing internal space $M_0$. 
The solutions are valid if the dimensions of $p$-branes and dilatonic 
coupling vector satisfy the orthogonality restrictions (\ref{5.16}).

The special case $w=+1$, $M_0 = S^{d_0}$ corresponds
to spherically-symmetric configurations containing black
hole solutions (see, for example \cite{CT,AIV,O}).  It may
be interesting to apply the relations from Sect. 6 to
minisuperspace quantization of black hole configurations
in string models, M-theory etc.

\end{document}